\documentclass[natbib]{article}
\usepackage[utf8]{inputenc} 
\usepackage[T1]{fontenc}    
\usepackage{hyperref}       
\usepackage{url}            
\usepackage{booktabs}       
\usepackage{amsfonts}       
\usepackage{nicefrac}       
\usepackage{microtype}      
\usepackage{lipsum}
\usepackage{amsmath}
\usepackage{bbm}
\usepackage{subcaption}
\usepackage{graphicx}
\usepackage{mathptmx}
\usepackage{verbatim}
\usepackage{natbib}
\usepackage[ruled,vlined]{algorithm2e}
\SetKw{KwFrom}{from}

\usepackage{caption}
\begin{document}
\title{High-dimensional clustering via Random Projections}

\author{Laura Anderlucci\footnote{via delle Belle Arti 41, 40126 Bologna (Italy) \newline mailto: \url{laura.anderlucci@unibo.it}} \and Francesca Fortunato \and Angela Montanari}

\date{}
\maketitle
\vspace{-1cm}\begin{center} Department of Statistical Sciences -  University of Bologna
\end{center}

\vspace{1cm}
\begin{abstract}
In this work, we address the unsupervised classification issue by exploiting the general idea of Random Projection Ensemble. Specifically, we propose to generate a set of low dimensional independent random projections and to perform model-based clustering on each of them. The top $B^*$ projections, i.e. the projections which show the best grouping structure are then retained. The final partition is obtained by aggregating the clusters found in the projections via consensus. The performances of the method are assessed on both real and simulated datasets. The obtained results suggest that the proposal represents a promising tool for high-dimensional clustering.

\vspace{0.5cm}
\noindent%
\vfill
\end{abstract}


\section{Introduction}

Data clustering plays a key role in modern statistics as it represents one of the most effective tools to understand the underlying structure of a given data set. The aim of clustering is essentially to categorize data into `clusters' (or groups) such that observations belonging to the same cluster are more similar to each others than those in different groups. This problem has been studied extensively and the state-of-the-art is exposed in surveys that have appeared regularly over the years; see, for example, \cite{mclachlan2019finite}, \cite{bouveyron2014model}, \cite{xu2015comprehensive}.

Clustering in low-dimensional spaces requires limited resources; the complexity of the problem indeed increases with the number of observed features, $p$.
When dealing with high-dimensional data, the use of traditional unsupervised classification algorithms faces several limitations; in particular, the presence of noisy or irrelevant information can mislead these methods due to the `\emph{curse of dimensionality}', as coined by \cite{bellmandynamic}. In order to overcome this problem, often dimension reduction procedures are applied before carrying out any clustering.

Generally, the term `dimension reduction' refers to two different approaches; namely, it includes both \emph{feature selection} methods that embed the high-dimensional points into a lower subspace by selecting some `relevant' variables, and \emph{feature extraction} algorithms which find an embedding by constructing new artificial features that are, for example, linear combinations of the original ones.
Variable selection strategies have been frequently used to handle high-dimensional clustering issues, but feature extraction procedures could be generally more efficient. Feature selection techniques indeed may discard some potentially important variables, e.g. variables that are not predictive if individually considered, but that could provide significant benefits when taken in conjunction with other features.

Traditionally, variable combination methods involve the projection of high-dimensional data onto a lower subspace with the intent of capturing as much of the data variability as possible (e.g. Principal Component Analysis). Albeit this approach has been successfully used in many applications, its aim does not always coincide with that of a clustering task. In fact, the useful information about the group structure is not necessarily contained in the subspaces with the largest variance, as exposed by \cite{chang1983using}.
A recent approach for dimension reduction that has been gaining increasing attention is based on Random Projections (RPs) and consists in mapping at random the original high-dimensional data onto a lower subspace by using a random matrix with orthogonal columns of unit length. Specifically, the key point of RP is that, regardless of the original data dimension, the final solution still preserves the global information almost perfectly. Such a result is guaranteed by the \cite{johnson1984extensions} Lemma, which states that any $n$-point set in $p$ dimensions ($X = [\mathbf{x}_1, \dots, \mathbf{x}_i, \dots, \mathbf{x}_n]^T$, $\mathbf{x}_i \in \mathbb{R}^p$, $ i=1,\dots,n$) can be linearly projected onto $d=O(\log(n)/\epsilon^2)$ coordinates (with $d \ll p$) , by using a random matrix $A \in \mathbb{R}^{p\times d}$ with orthonormal columns, while preserving pairwise distances within a factor $1\pm \epsilon$. More precisely, with high probability over the randomness of $A$:

\begin{equation}\label{eq:JL}
(1-\epsilon)||\mathbf{x}_i-\mathbf{x}_j||_2 \leq ||A^{\top}\mathbf{x}_i-A^{\top}\mathbf{x}_j||_2 \leq (1+\epsilon)||\mathbf{x}_i-\mathbf{x}_j||_2,
\end{equation}

where $|| \cdot ||_2$ indicates the $L_2$ norm.

\cite{bhattacharya2009low} proved that also the Hellinger distance between any two distributions $P$ and $Q$, defined as
\begin{equation*}\label{eq:hellinger}
H(P,Q) =\frac{1}{2}\left(||\sqrt{P}- \sqrt{Q} ||_2\right)^2,
\end{equation*}

admits a low disorsion JL-type embedding.
In the model-based clustering context, where data are considered as coming from a distribution that is a mixture of two or more components, this theorem directly implies that the distance between the density of any pair of components is preserved with arbitrarily small distortion. In other words, it states that if two component densities are sufficiently far apart in the high-dimensional space, then they would be expected approximately the same also in the reduced $d$-dimensional space.

\medskip
These interesting results motivated us to employ random projections within a model-based clustering framework. Specifically, inspired by the original idea of \cite{cannings2017random} for supervised classification, we propose to generate a set of $B$ low dimensional independent random projections and to apply a Gaussian Mixture Model (GMM) on each of them. Our Random Projection Ensemble Clustering (RPE Clu) algorithm then obtains the final partition by combining via consensus the clustering results from the top $B^*$ projections, i.e. the projections which show the best grouping structure according to a given criterion.

\medskip

The paper is organized as follows. Section 2 recalls the model-based clustering framework. In the same section, some popular dimension reduction procedures for high-dimensional clustering are briefly presented. In Section 3, the Random Projection Ensemble Clustering algorithm (RPE Clu) is introduced and defined in detail. Section 4 is devoted to practical considerations about the computational complexity of the algorithm, the choice of the number of random projections and the dimension of the projected space. Section 5 presents a simulation study where the proposed methodology is compared with some benchmark clustering techniques. In Section 6, RPE Clu is applied to two sets of high-dimensional real data. A final discussion on the obtained results concludes the paper.

\section{High-dimensional model-based clustering} \label{HDMBC}
In model-based clustering (see \cite{peel2000} for a detailed review), data are assumed to derive from a common source with $G$ different sub-populations. In particular, each sub-population is modelled separately (typically by members of the same parametric density family) and the overall population is but a mixture of them. The resulting model is a finite mixture and it is described by the following probability density function (pdf):

\[
  f(\mathbf{x})=\sum_{k=1}^G \pi_k f_k(\mathbf{x}| \mathbf{\theta}_k).
\]

Here, $f_k$ and $\mathbf{\theta}_k$ are the density and the parameters of the $k$-th component of the mixture, respectively, whereas $\pi_k$ is the prior probability that an observation belongs to the $k$-th component ($\pi_k \geq 0$, $\sum_{k=1}^G \pi_k = 1$). For clustering purposes, units are allocated to the component whose posterior probability is maximum.

A common choice for $f_k(\cdot)$ is the multivariate normal distribution, $\phi_k(\cdot)$, parameterized by its mean vector $\mathbf{\mu}_k$ and its covariance matrix $\Sigma_k$:

\[
  \phi_k(\mathbf{x}|\mathbf{\mu}_k, \Sigma_k) = (2\pi)^{-(p/2)} |\Sigma_k|^{-(1/2)} \exp\left\{-\frac{1}{2} (\mathbf{x}-\mathbf{\mu}_k)'\Sigma_k^{-1}(\mathbf{x}-\mathbf{\mu}_k) \right\}.
\]

Following this approach, the entire data set is modeled by a Gaussian Mixture model:

\[
  f(\mathbf{x})=\sum_{k=1}^G \pi_k \phi_k(\mathbf{x}|\mathbf{\mu}_k, \Sigma_k).
\]

\medskip

In presence of high-dimensional data the GMM tends to perform poorly, due to the large number of parameters to estimate with relatively few observations. In fact, the number of parameters increases quadratically with $p$ and thus the maximum-likelihood estimation problem becomes ill-posed very quickly. The earliest approaches which appeared in the literature to overcome this limit and attain parsimony propose alternative parameterizations of the component densities. For instance, \cite{banfield1993model} and \cite{celeux1995gaussian} introduce a parsimonious parameterizations of the covariance matrix in terms of its eigenvalue decomposition so as to control the volume, shape and orientation of the Gaussian ellipsoids. \cite{biernacki2014stable} define different parsimonious models based on a variance-correlation decomposition of the covariance matrices.

When performing variable selection for clustering, the aim is essentially to identify those features that bring relevant information about the underlying group structure. In the model-based context, the definition of `relevance' should be expressed in terms of probabilistic dependence (or independence) with respect to $\mathbf{Z}$, i.e. the random vector  which describes the latent class membership ($\mathbf{Z} = [Z_1, \dots, Z_k, \dots, Z_G]^T$, $Z_k \in \{0,1\}$, $k=1,\ldots , G$). Specifically, the distribution of relevant variables directly depends on $\mathbf{Z}$ as these features contain the key clustering information. Conversely, both redundant and uninformative variables do not provide any additional or useful information and, thus, they can be assumed to be conditionally independent given the relevant variables or completely independent of the group structure, respectively.
Following this approach, several authors have recast the variable selection problem for clustering in a model selection one. Namely, relevant variables are sought through a stepwise procedure that, at each step, compares models that differ in the role assigned to the variables in explaining the clustering structure.

Pioneers of this framework were \cite{raftery2006variable}, who introduced a procedure in which the decision for inclusion or exclusion of a generic (set of) variable(s) $\mathbf{x}^P$ into the current set of clustering ones $\mathbf{x}^C$ is taken by comparing two competing models in terms of their Bayesian Information Criterion (BIC). In particular, Model I assumes that $\mathbf{x}^P$ carries relevant information about the cluster membership, whereas Model II states that $\mathbf{x}^P$ does not depend on $\mathbf{Z}$. The BIC associated to these models are:

\begin{align} \label{eq:BIC_varsel}
\text{BIC}_I &= \text{BIC}_{\text{clust} }(\mathbf{x}^C, \mathbf{x}^P) \nonumber \\
\text{BIC}_{II} &= \text{BIC}_{\text{clust} }(\mathbf{x}^C) + \text{BIC}_{\text{reg}}(\mathbf{x}^P|\mathbf{x}^C)
\end{align}	

Here, BIC$_{\text{clust} }(\mathbf{x}^C, \mathbf{x}^P)$ is the BIC of the GMM in which $\mathbf{x}^P$ adds useful information, BIC$_{\text{clust} }(\mathbf{x}^C)$ is the BIC of the GMM on the current set of clustering variables only and BIC$_{\text{reg}}(\mathbf{x}^P|\mathbf{x}^C)$ is the BIC of the regression of $\mathbf{x}^P$ on $\mathbf{x}^C$.
If  BIC$_I-$ BIC$_{II} >0$, then $\mathbf{x}^P$ is added to the set of clustering variables $\mathbf{x}^C$.

This method has been further improved by \cite{maugis2009variablea} and \cite{maugis2009variableb} under the assumption that the irrelevant variables can be independent of some relevant ones.

\medskip
Recently, two further extensions of the above modeling appeared in the literature: \cite{scrucca2016genetic} suggests to overcome the sub-optimality of a stepwise model search by employing genetic algorithms; \cite{galimberti2018modelling} take into account the possibility that different variable vectors provide information about different clustering structures.

\medskip
Although effective in many applications, in the unsupervised classification context the variable selection problem is ill-posed: clusters indeed strongly depend on the selected features and the features are selected according to the clusters (see \cite{ruiz2009information}). For this reason, feature extraction procedures would rather be preferred.


\cite{dasgupta2013experiments} demonstrated that RPs can be successfully used to handle high-dimensional clustering issues with a model-based approach. Firstly, he showed that a mixture of $G$ Gaussians can be embedded onto just $O(\log G)$ random coordinates without destroying the original group structure. Second, he proved that even if the original Gaussians exhibit eccentric elliptical contours, their projected counterparts are always more spherical. These two benefits are of major importance and they definitely facilitate the learning of a Gaussian Mixture Model. In particular, dimension reduction saves a lot of time and computational costs on one hand; on the other, clusters of low eccentricity reduce the EM algorithmic challenges ensuring that intermediate covariance matrices are not singular or close to singular.

\section{Random projection ensemble clustering}
As discussed in the previous section, high-dimensional data pose many challenges to model-based clustering. Methods in this class indeed become rapidly over-parameterized since the number of parameters to estimate increases quadratically with the number of observed features $p$.

Random projections have shown to provide promising results for the analysis of high-dimensional data. Their main inconvenience is that they are highly unstable: namely, different random projections of the original data may provide completely different classification results. That is the reason why most of the successful proposals on RPs resort to ensembles. For example, \cite{fern2003random} propose to aggregate the clustering results of a GMM on different random projections of the data into a similarity matrix containing the probability ``estimates'' that any two data points belong to the same cluster;  then, they suggest to perform an agglomerative clustering procedure on such a matrix to produce the final groups.

In this paper, we also exploit the general idea of RP ensemble for high-dimensional clustering. In particular, our novel proposal consists of applying a Gaussian Mixture Model to \emph{carefully chosen} random projections of the original data, but differently from Fern and Broadley, we use the GMM properties for both projection selection and consensus aggregation.

\subsection{On the choice of random projections}
Differently from other transformation techniques (such as, for example, principal components or projection pursuit), the random projection method does not exploit any `interestingness' criterion to identify the `optimal' projection. High-dimensional data are just embedded into a lower dimensional subspace by using a random projection matrix $A$ with orthogonal and unit length columns \citep[e.g. generated according to Haar measure,][]{Haar}.  As a consequence of that, results from distinct configurations of the same data can be even dramatically different: some projections indeed can highlight a clear group structure in the lowered data, whilst some others can derail any hope of learning by confusing all the groups together.

In this section, we propose a method for choosing a number $B^*$ of `good' random projections, that is, a criterion for identifying those projections showing a clear group structure.

\cite{hennig2019cluster} provides a detailed review of the validation indexes proposed in the literature to evaluate the quality of a clustering procedure. Although effective, many of these indexes rely on a measure of distance/dissimilarity and, therefore, they may seem inconsistent with a model-based framework.
Furthermore, since in the unsupervised context no \textit{apriori} information about the structure being looked for is available, we believe it makes sense to consider the RP selection as a part of the clustering algorithm, i.e. as the choice of the model that best fits the data according to a specific criterion (e.g. the BIC).

The BICs of mixture models fitted to different random projections cannot in principle be compared, because they are referred to different variables generated by the different random projections. On the contrary, the BIC of different models defined in the original variable space can be compared.
We search for the solution that maximizes the log-likelihood of the GMM fitted on the original data, penalized by the number of free parameters.

In practice, in order to avoid the drawbacks associated with the high-dimensional spaces, a feasible solution consists in considering the following variable partition

\[ Y^* = [Y, \bar{Y}] =  [XA | X\bar{A}],\]

where $X \in \mathbb{R}^{n \times p}$ is the original high-dimensional data matrix, $A \in \mathbb{R}^{p\times d}$ is the random projection matrix and $\bar{A} \in \mathbb{R}^{p \times (p-d)}$ is its orthogonal complement.
The basic idea is to perform model-based clustering on the reduced data $Y = XA$, assuming that the underlying group structure may be well approximated by the one in the $d$ dimensions of the block matrix $Y^*$, i.e.:
\begin{equation} \label{eq:cond_dens}
	f(Y|\mathbf{Z}=\mathbf{z})=\sum_{k=1}^G \pi_k \phi_k(Y|\mathbf{\mu}_{Y_k}, \Sigma_{Y_k}).
\end{equation}
This assumption does not imply that $\bar{Y}$ is not useful for clustering, but only that it contains some information on the group membership $\mathbf{z}$ that is very similar to that already available in $Y$. Therefore, in terms of distributional representation, it seems reasonable to think of $\bar{Y}$ as conditionally independent of $\mathbf{Z}$ given $Y$; it could be necessary for the clustering, but only if $Y$ is not present \citep{fop2018variable}. This amounts to assume that:
\begin{equation} \label{eq:cond_dens}
  f(\bar{Y}|Y) = \phi(\bar{Y}|\mathbf{\mu}_{\bar{Y}|Y}, \Sigma_{\bar{Y}|Y})
\end{equation}
where
\begin{align} \label{eq:schur}
\mathbf{\mu}_{\bar{Y}|Y}&= \mathbf{\mu}_{\bar{Y}} + \Sigma_{\bar{Y}Y} \Sigma_{Y}^{-1}(Y-\mathbf{\mu}_Y) \nonumber \\
\Sigma_{\bar{Y}|Y}&=\Sigma_{\bar{Y}} - \Sigma_{\bar{Y}Y} \Sigma_{Y}^{-1} \Sigma_{Y\bar{Y}}.
\end{align}
Equation (\ref{eq:schur}) describes the Schur complemement of the block $\Sigma_{Y}$ in the $p \times p$ block-matrix

\[
\Sigma_{Y^*}=
\left[
\begin{array}{ll}
\Sigma_{Y} & \Sigma_{Y\bar{Y}} \\
\Sigma_{\bar{Y}Y} & \Sigma_{\bar{Y}}
\end{array}
\right].
\]

The distribution of $Y^*$ is the product of the marginal density of $Y$, $f(Y|\mathbf{z})$, and the conditional density of $\bar{Y}|Y$, $f(\bar{Y}|Y)$:
\begin{equation} \label{eq:joint_dens}
	f(Y^*|\mathbf{Z}=\mathbf{z}) = \left[\sum_{k=1}^G \pi_k \phi_k(Y|\mathbf{\mu}_{Y_k}, \Sigma_{Y_k})\right]\phi(\bar{Y}|\mathbf{\mu}_{\bar{Y}|Y}, \Sigma_{\bar{Y}|Y})= \sum_{k=1}^G \pi_k \phi_k(Y^*|\mathbf{\mu}_{Y^*_k}, \Sigma_{Y^*_k}).
\end{equation}

%
%
%
%

Equation (\ref{eq:joint_dens}) can be easily rewritten in terms of log-likelihood as:

\begin{equation} \label{eq:joint_loglik}
\sum_{i=1}^{n}\log[f(\mathbf{y}_i^*|\mathbf{Z}_i=\mathbf{z}_i)] = \sum_{i=1}^{n} \log[f(\mathbf{y}_i)|\mathbf{Z}_i=\mathbf{z}_i] + \sum_{i=1}^{n} \log[f(\bar{\mathbf{y}
}_i|\mathbf{y}_i)].
\end{equation}

\medskip
The BIC corresponding to Equation (\ref{eq:joint_loglik}) is:

\begin{equation} \label{eq:BIC}
\text{BIC} = \text{BIC}_{\text{GMM}}(Y) + \text{BIC}_{\text{reg}}(\bar{Y}|Y),
\end{equation}
where $\text{BIC}_{\text{GMM}}(Y) = 2\log[f(Y)] - q_Y \log(n)$ is the BIC associated to the Gaussian mixture fitted on the $d$-dimensional data and $\text{BIC}_{\text{reg}}(\bar{Y}|Y) = 2\log[f(\bar{Y}|Y)] - q_{\bar{Y}} \log(n)$ is the BIC for the linear regression of the $(p-d)$ last columns of $Y^*$ on the first $d$ ones. In high dimensional clustering the dimension of $Y$ is generally much smaller than that of $\bar{Y}$ and is the same for each projection; therefore, the size of $Y$ only slightly affects the penalty term in the $\text{BIC}_{\text{reg}}$. This condition does not hold in the scenarios described in \cite{maugis2009variablea} and \cite{clustvarsel}.

The number of free parameters of the GMM on $Y$ and those of the linear regression are described by $q_Y$ and $q_{\bar{Y}}$, respectively.
In order to allow for great flexibility, $\Sigma_{\bar{Y}|Y}$ is assumed to have a general form and, thus,
\[
q_{\bar{Y}} = (p-d)(d+1)+\frac{(p-d)[(p-d)+1)]}{2}.
\]

When the number of observed features $p$ is particularly large with respect to $d$, a restricted form for $\Sigma_{\bar{Y}|Y}$ is suggested. Namely, $\Sigma_{\bar{Y}|Y} = \text{diag}(\sigma_1^2, \dots ,\sigma_{p-d}^2)$. In this case, the number of free parameters for the regression model reduces to $q_{\bar{Y}} = (p-d)(d+1)+(p-d)$.

\begin{figure}
  \centering
 \includegraphics[width=.8\textwidth]{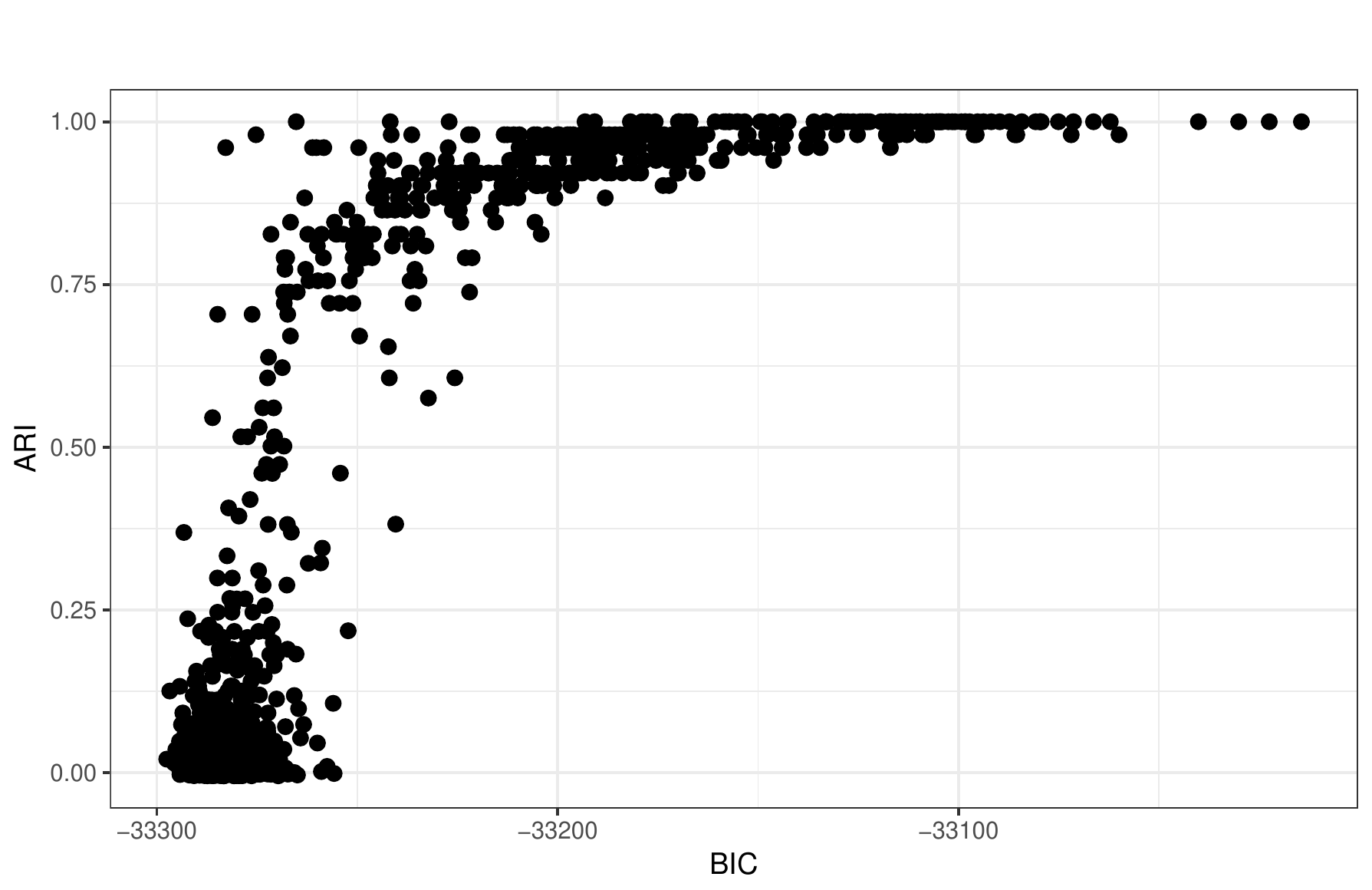}
  \caption{ARI of the classification yielded by the GMM on $B=1000$ different $8$-dimensional projections of a simulated dataset with $p=100$, $G=2$, $n_1=n_2=100$ and the true class membership, ordered by increasing values of the BIC.}\label{fig:Ari_Bic}
\end{figure}

\medskip
As depicted in Figure \ref{fig:Ari_Bic}, the criterion we propose provides a good ranking of the random projections according to the goodness of the partition they induce. Specifically, models with increasing BIC exhibit large values for the Adjusted Rand Index (ARI), i.e. a measure of the similarity between the classification yielded by the GMM on the reduced data and the true class membership.

Such result does not imply that the projections associated with the largest BIC would yield redundant solutions. In fact, the random projection method naturally perturbs different configurations of the original data, thus, inducing diversity.
Pairwise Adjusted Rand Indexes on all clustering vectors returned by the selected top 100 projections largely differ. For example, in the scenario depicted in Figure \ref{fig:Ari_Bic} the average ARI value ranges from 0.6170  to  0.9845, for 100 replications.


\subsection{On the result aggregation}
A possible solution to the inherent instability associated with random projections involves the use of cluster ensembles that combine multiple individual partitions into a single consensus one. This process was pioneered by \cite{strehl2002cluster} who proved that ensembles can provide robust and stable solutions across different problem domains. A detailed review of the state-of-the-art cluster ensemble methods can be found in \cite{boongoen2018cluster}, where both theoretical aspects and empirical applications are widely discussed.

Consensus clustering algorithms generally derive the ultimate data partition by minimizing an objective function that measures how dissimilar each hard or soft consensus candidate is from the ensemble members. In this work, we suggest to derive the final unit allocation by using the greedy algorithm proposed by \cite{dimitriadou2002combination} and developed in \cite{hornik2005clue};  in the following, a description of their procedure is sketched.

The aim is to look for a partition $P$ of the given dataset $\{x_1,\ldots,x_n\}$ into $G$ classes that optimally represents a given set of $B^*$ partitions of the same set. Each of these $B^*$ partitions is represented by an $n \times G$ membership matrix $U^{(b)}$, $b=1,\ldots,B^*$. The element $u_{ik}^{(b)}$ of $U^{(b)}$ is the membership of $x_i$ to the $k$th class, $k=1,\ldots,G$, of the $b$th partition. The final partition $P$ is encoded as an $n \times G$ matrix with element $p_{ik}$. In order for the partition $P$ to be optimal, it needs to be at the smallest distance from the considered $B^*$ partitions. Thus, the task is to find $P$ in such a way that
\[ \min_{P} \left(\frac{1}{B^*} \sum_{b=1}^{B^*} h(U^{(b)},P) \right),\]
where $h(U^{(b)},P)$ is the dissimilarity function between $U^{(b)}$ and $P$:
\[ h\left(U^{(b)},P\right) =\frac{1}{n} \sum_{i=1}^n \left\lVert \mathbf{u}_i^{(b)} - \mathbf{p}_i \right\rVert^2.\]

However, because of the label switching clustering issue, any relabeling of the classes is to be considered as the same partition. Thus, partitions $U^{(b)}$ and $\Pi_b(U^{(b)})$,  which only differ by a permutation of the class labels are to be considered the same and the distances should remain the same too:
\[ h\left(U^{(b)}, P\right) = h\left(\Pi_b (U^{(b)}),P \right), \quad \forall P.\]

Therefore, the dissimilarity function $ h(U^{(b)},P)$ between two clustering partition $U^{(b)}$ and $P$ should be rather defined as:
\[ h\left(U^{(b)},P\right) =\min_{\mathbf{\pi}_b} \left( \frac{1}{n} \sum_{i=1}^n \left\lVert \mathbf{\pi}_b(\mathbf{u}_i^{(b)}) - \mathbf{p}_i \right\rVert^2 \right).\]
where the minimum is taken over all possible column permutations $\pi_b$.

The task of finding an optimal partition $P$ is then given by the minimization problem
\[ \min_{p_1,\ldots,p_n} \min_{\mathbf{\pi}_1,\ldots, \mathbf{\pi}_{B^*}} \left( \frac{1}{B^*} \sum_{b=1}^{B^*} \frac{1}{n} \left\lVert \mathbf{\pi}_b(\mathbf{u}_i^{(b)}) - \mathbf{p}_i \right\rVert^2 \right).\]
In order to find an optimal $P$, $\mathbf{p}_i$ and $\mathbf{\pi}_b$ have to be minimized simultaneously, because the choice of the permutations $\mathbf{\pi}_b$ depends on $\mathbf{p}_i$. As a direct solution of the minimization problem is unfeasible, a greedy algorithm is employed.

The iterative procedure determines, at each step $b$ ($b=1,\ldots B^*$), the locally optimal permutation matrix $\Pi_b$ for relabeling by minimizing the Euclidean distance between the previously determined consensus candidate $P^{(b-1)}$ (note that at the initial step $P^{(0)} \equiv U^{(1)}$) and all the possible permutations of the membership matrix $U^{(b)}$, $\Pi_b U^{(b)}$. Then, it derives the updated consensus partition by:
\[ P^{(b)} =\frac{1}{b} \sum_{l=1}^b \hat{\Pi}_l(U^{(l)})=\frac{b-1}{b} P^{(b-1)}+\frac{1}{b} \hat{\Pi}_b(U^{(b)}).\]
 In so doing, this sequential method helps to tackle the issue of simultaneous combination of all partitions, otherwise computationally unfeasible. For further details, see \cite{dimitriadou2002combination}, and \cite{hornik2005clue} for the corresponding R package.

\subsection{Random projection ensemble clustering algorithm}
In this paper, a new model-based clustering method for high-dimensional data based on random projections, is introduced. The algorithm is sketched in the following:

\medskip
\begin{algorithm}[H] \label{alg:RPEclu}
\caption{RPE Clu}
\SetAlgoLined
\SetAlgoHangIndent{1cm}

\KwResult{Partition of the original data $X$ into $G$ groups.}
Set $G$, $d$, $B$, $B^*$\;
1. \For{ $b$ \KwFrom 1 \KwTo $B$}
 {
$(i)$ Generate an independent $d$-dimensional random projection matrix $A_b$\;
$(ii)$ Fit a GMM with $G$ components on the projected data $Y=XA_b$\;
$(iii)$ Retain the induced data partition $C_b$\;
$(iv)$ Fit linear regression of $\bar{Y}$ on $Y$, where $\bar{Y}=X\bar{A}_b$ and $\bar{A}_b$ is the orthogonal complement of $A_b$\;
$(v)$ Compute the BIC as described in Equation \ref{eq:BIC} \;
}

2. Sort the BIC values of the $B$ solutions\;
3. Select the top $B^*$ projections from the BIC list of point 2\;
4. Aggregate the corresponding $B^*$ cluster membership vectors via consensus\;
5. Partition the original data $X$ according to the consensus membership of point 4\;
\end{algorithm}

\section{Practical considerations}
\subsection{Computational complexity}
The algorithm we propose derives the final partition by aggregating the results of Gaussian Mixture Model clustering performed on an `optimal' subset of random projections.

The first step of this procedure involves the computation of $B$ random projection matrices. The cost of this operation varies according to the method used: namely, generating a single RP from the Haar measure requires $O(pd^2)$ operations, whilst choosing each entry of this matrix uniformly and independently from $[-1,1]$ takes time only $O(dp)$ \citep[see][]{achlioptas2003database}.

Once the projections have been generated, the original high-dimensional data should be embedded onto the lowered spaces; each projection requires $O(npd)$ operations.

Then, for $b = 1, \dots, B$, a GMM is performed on the reduced set $Y = XA_b$ with a total cost of $O(Gd^3)\cong O(d^3)$. Simultaneously, a multiple linear regression of $\bar{Y}=X\bar{A}_b$ on $Y$ is computed. The cost of this step is $O((p-d)^3)$. Finally, the BIC values computed as in Equation (\ref{eq:BIC}) are sorted and observations are clustered by using the best $B^*$ projections (i.e. those yielding the highest values for the BIC). These steps involve $O(B^*)$\footnote{See the R Documentation for the \texttt{sort} function with default settings.} and $O(B^*nd)$ resources, respectively.

\subsection{Choice of $B$ and $B^*$}
The random projection ensemble clustering performances strongly depend on the possibility to identify those random projections that induce a very clear group structure in the reduced space.

The choice of $B^*$, i.e. the number of `base' models to retain in the final ensemble, is more insidious.
Several studies have shown that ensembles of classifiers are generally more effective when they are constructed from members whose errors are dissimilar; see, for example, \cite{kittler1998combining}. In fact, aggregating the base results of models that agree on how a dataset should be partitioned does not provide any improvement.
The random projection method itself represents a valid technique to introduce artificial instability (and thus \textit{diversity}) to an ensemble as it allows to generate clustering results from different perturbed configurations of the original data. However, as \cite{fern2003random} point out, taking into account too many projections may degrade the final result, especially when the original features are highly correlated; furthermore, it surely increases the computational cost of the procedure. On the other side, considering a very small ensemble can be risky, too. In fact, since in clustering no \textit{apriori} knowledge of the true data structure is available, identifying the best predictors is not a trivial task and, therefore, any criterion (including the BIC we propose) could be confused. In order to avoid the selection of too similar or inaccurate base classifiers, a compromise solution for $B^*$ is highly suggested.

On the basis of the numerical evidences we suggest $B=1000$ and $B^*=100$ as generally good choices.



\subsection{Choice of $d$}
\cite{dasgupta2013experiments} proved that data from an arbitrary mixture of $G$ Gaussian distributions can be randomly embedded into a subspace of just $O(\log G)$ dimensions, while preserving the group structure almost perfectly. Furthermore, if $d<\log G$, the worsening of the mapping performance is gradual. This result is particularly appealing as it proves that the dimension of the projection subspace is independent of the original dimensions of the data, that is, $d$ does not depend upon $n$ nor $p$.
A couple of numerical experiments conducted on both simulated and real data, corroborate Dasgupta's result.

\begin{figure}
  \centering
  \includegraphics[width=\textwidth]{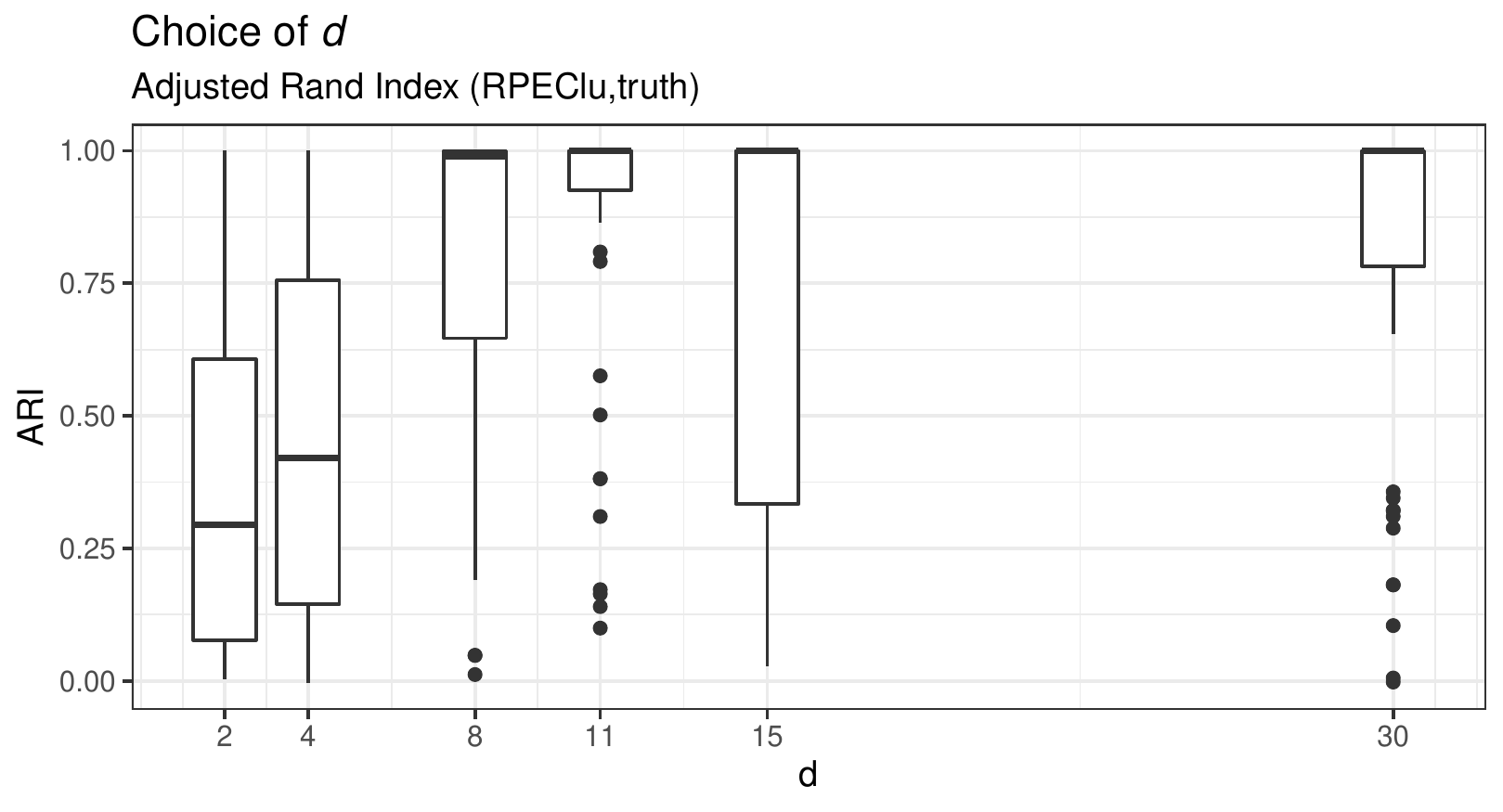}
  \caption{Study on the choice of $d$.  The boxplots report the ARI of the random projection ensemble clustering algorithm for different dimensions of the projected space, $d$. The numbers of generated and selected projections are set equal to $B=1000$ and $B^*=100$, respectively.}\label{fig:sc2_varying_d}
\end{figure}

In particular, we generated data from two $p=500$-dimensional Gaussian populations, with correlated features ($0.90$ for all of the features); the size of each sample was set to 100. RPE Clu was applied for different values of $d$ (namely, $\{2,4,8,11,15,30\}$, corresponding to  $[1 \log(G)]+1, [5 \log(G)]+1,[10 \log(G)]+1,[15 \log(G)]+1,[20 \log(G)]+1,[42 \log(G)]+1$), with $G$ considered as known (and equal to 2), $B=1000$, $B^*=100$ and no contraints were imposed on the covariance matrices; 50 replications were performed. Figures \ref{fig:sc2_varying_d} reports the Adjusted Rand Index of the obtained partition and the true class membership for different $d$ values.

\begin{figure}
  \centering
  \includegraphics[width=.8\textwidth]{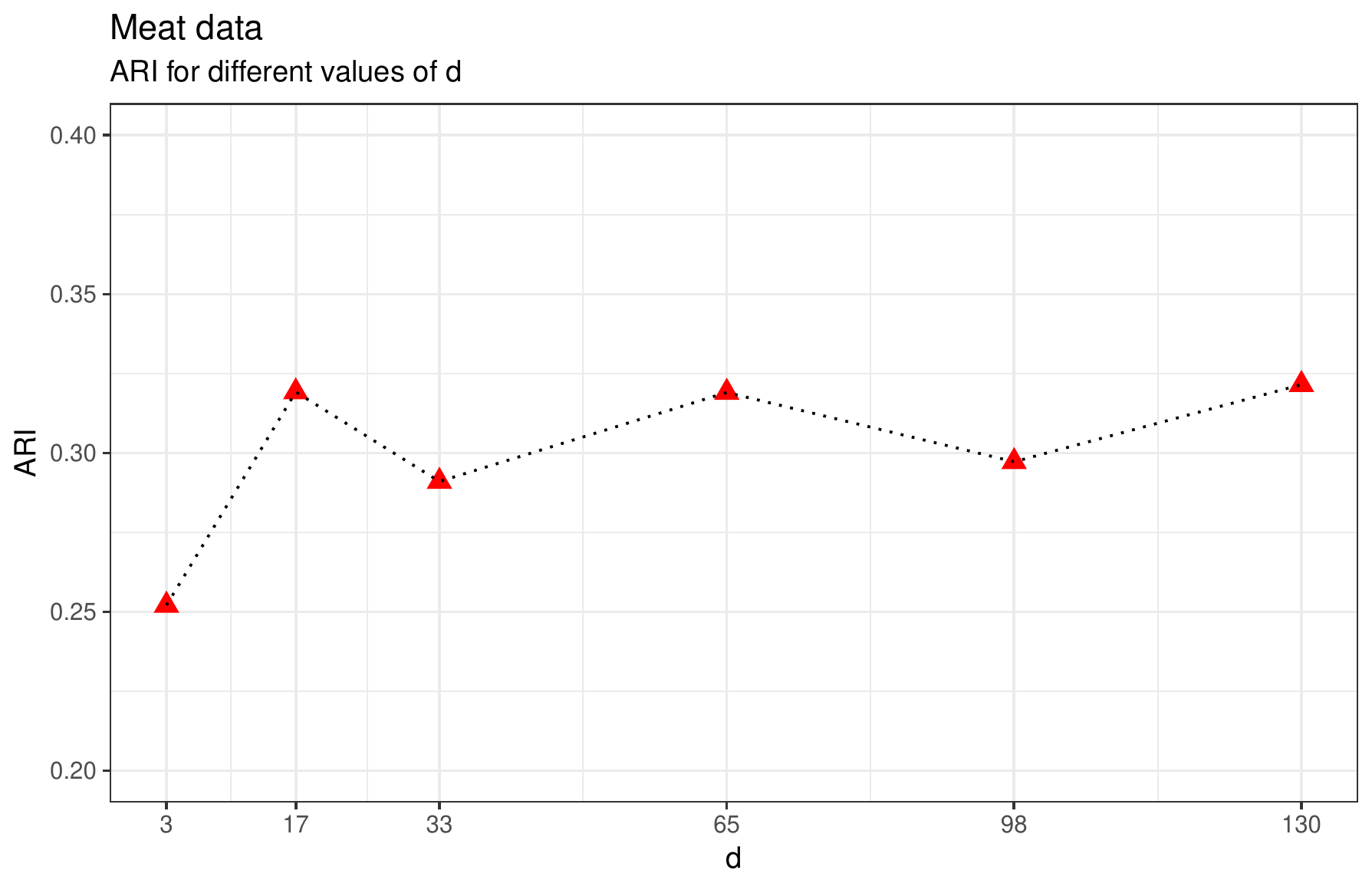}
  \caption{Meat data. ARI of the random projection ensemble clustering algorithm for different dimensions of the projected space, $d$. The numbers of generated and selected projections are set equal to $B=1000$ and $B^*=100$, respectively.}\label{fig:meat_varying_d}
\end{figure}

The real data (fully described in Section \ref{subsec:meat data}) refer to $n=231$ samples  of homogenized raw meat, whose spectra are recorded; the total number of variables is $p=1050$ and $G=5$. RPE Clu is run with $B=1000$, $B^*=100$ and different values of $d=\{3,17, 33,65,98,130\}$ corresponding to $[1 \log(G)]+1, [10 \log(G)]+1,[20 \log(G)]+1,[40 \log(G)]+1,[60 \log(G)]+1,[80 \log(G)]+1$; $G$ was considered as known and equal to 5, no constraints were imposed on the covariance matrices. Figures \ref{fig:meat_varying_d} reports the Adjusted Rand Index of the obtained partition and the true class membership for different $d$ values.

Figures \ref{fig:sc2_varying_d} and \ref{fig:meat_varying_d} clearly show that a choice of $d=[10 \log(G)]+1$ works pretty well; higher values of $d$ do dot noticeably improve the final performance.

\section{Simulation study}
The performance of the RPE Clu algorithm is evaluated in a variety of scenarios through an extensive simulation study. In particular, $G = \{2,4\}$ different Gaussian clusters of size 100 are generated in $p = \{100, 500, 1000\}$ dimensions by using the \verb"sim_normal" function of the \verb"clusteval" \verb"R" package \citep{clusteval}. According to its parametrization, each population has a $p$-dimensional multivariate normal distribution, with mean vector
\[\mathbf{\mu}_k = \frac{1}{2} \sum_{j=1}^{p/G} \mathbf{e}_{(p/G)(k-1)+j}, \]
where $\mathbf{e}_k$ is the $k$-th basis vector; therefore, the first $p/G$ dimensions of $\mathbf{\mu}_1$ are set to 1 and all the remaining to 0, the second $p/G$ dimensions of $\mathbf{\mu}_2$ are set to 1 and all the remaining to 0, and so on. The $k$-th population covariance matrix is
\[ \Sigma_k=(1-\tau_k)\mathbbm{1}_p+\tau_k{I}_p, \]
where $\mathbbm{1}_p$ and ${I}_p$ denote the $p \times p$ matrix of ones and identity matrix, respectively.
Here, $-(p-1)^{-1}<\tau_k<1$ governs the intra-class correlation; throughout the study, we evaluate different levels of correlation between variables, i.e. we take $\tau_k = \{0.1,0.3,0.4,0.6 \}$, corresponding to correlation values of $\{0.9,0.7,0.6,0.4 \}$ so as to explore how the clustering algorithm behaves in different situations.

As an exemplification, consider the simple case of $p=6$, $G=3$ and $\tau_k=0.1 \ \forall k$. The parameters of the three Gaussians are the following:

\begin{minipage}{0.4\textwidth}
\[ \mathbf{\mu}_1^{\top}=  \left(
                      \begin{array}{cccccc}
                        1 & 1 & 0 & 0 & 0 & 0 \\
                      \end{array}
                    \right)\]
\[ \mathbf{\mu}_2^{\top}=  \left(
                      \begin{array}{cccccc}
                        0 & 0 & 1 & 1 & 0 & 0 \\
                      \end{array}
                    \right)\]
\[ \mathbf{\mu}_3^{\top}=  \left(
                      \begin{array}{cccccc}
                        0 & 0 & 0 & 0 & 1 & 1 \\
                      \end{array}
                    \right)\]
\end{minipage} \quad
\begin{minipage}{0.5\textwidth}
\[ \Sigma_k=\Sigma=\left(
                     \begin{array}{cccccc}
                       1\ & 0.9\ & 0.9\ & 0.9\ & 0.9\ & 0.9 \\
                       0.9\ & 1\ & 0.9\ & 0.9\ & 0.9\ & 0.9 \\
                       0.9\ & 0.9\ & 1\ & 0.9\ & 0.9\ & 0.9 \\
                       0.9\ & 0.9\ & 0.9\ & 1\ & 0.9\ & 0.9 \\
                       0.9\ & 0.9\ & 0.9\ & 0.9\ & 1\ & 0.9 \\
                       0.9\ & 0.9\ & 0.9\ & 0.9\ & 0.9\ & 1 \\
                     \end{array}
                   \right)\]
\end{minipage}

\medskip

\begin{figure}
  \centering
  \includegraphics[width=\textwidth]{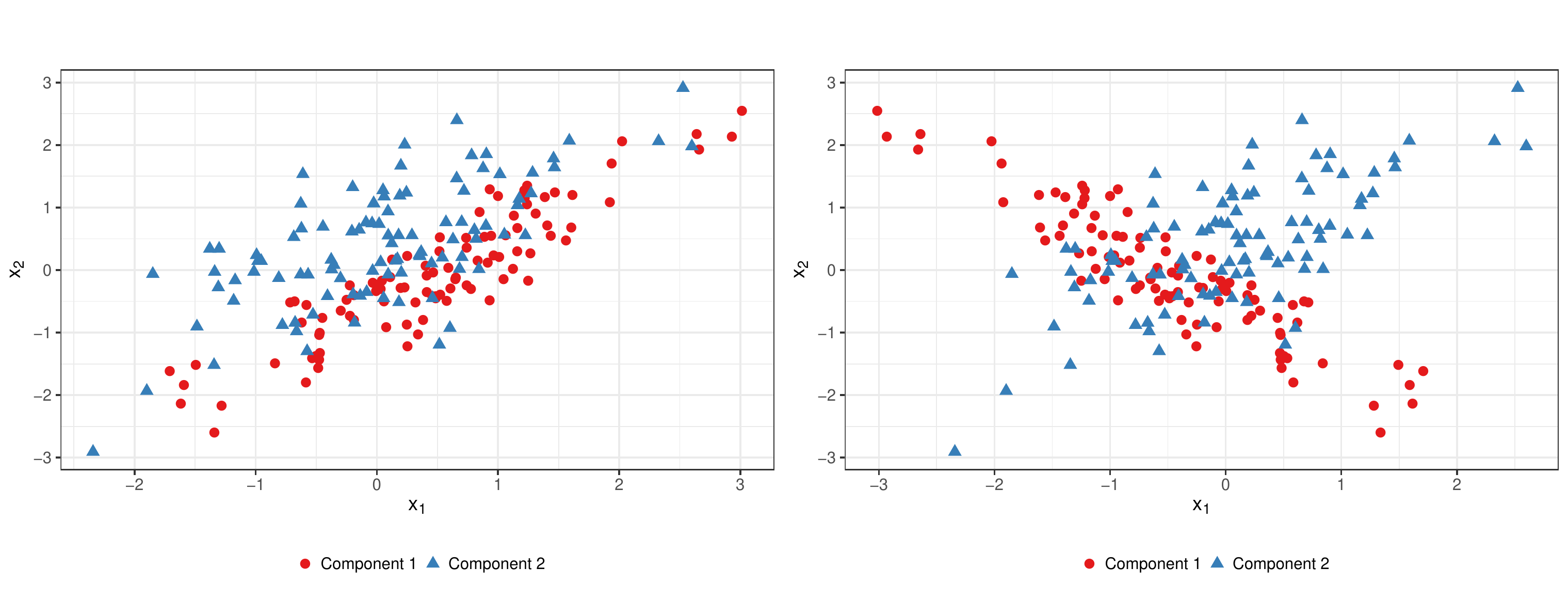}
  \caption{An example of bivariate dataset with \textit{heteroscedastic rotated} components: variable $x_1$ of Component 1 (red points) is rotated with respect to the axis $x=0$.}\label{fig:rotated_components}
\end{figure}

Furthermore, we consider scenarios characterized by both \textit{homoscedastic} (settings 1--12) and \textit{heteroscedastic} (settings 13--16) components. Scenarios with \textit{heteroscedastic rotated} components are also investigated (settings 17--20). In this case, as depicted in the illustrative example of Figure \ref{fig:rotated_components}, the first fifty odd variables of half of the groups are rotated with respect to the axis $x=0$.

In addition, we studied the behaviour of our proposal in contexts where original data deviate from Normality. In particular, settings 21--23 consider the exponential, the logarithm and the square-root transformation of $p$-variate Gaussian distributions, respectively ($p$=100, $n_k$=100, $k=1,\ldots,G$); the number of groups is set to two and only 50\% of the variables are relevant for clustering. Scenarios 24-26 extend the study to the case of four groups. A brief description of the simulation settings considered for the analysis is given in Table \ref{tab:settings} for Gaussian scenarios 1 - 20 and in Table \ref{tab:settings2} for non-Gaussian scenarios 21 - 26; more details are given in the Supplementary Material.

To validate the proposal, we apply other clustering algorithms on the same settings: the `standard' Gaussian Mixture Model \citep{peel2000} (via \verb"Mclust" function of the \verb"mclust" package), the $K$-means algorithm \citep{lloyd1982least} (via \verb"kmeans" function), Ward's agglomerative hierarchical clustering \citep{ward1963hierarchical} (via \verb"hclust" function) and the Partition Around Medoids (pam) \citep{kaufman2009finding} (via \verb"pam" function of the \verb"cluster" package). Two recent procedures that have shown good performances in the context of high-dimensional unsupervised classification are also included: namely, the Spectral clustering approach \citep{ng2002spectral} (\verb"specc" function of the \verb"kernlab" package) and the Affinity Propagation algorithm \citep{frey2007clustering} (\verb"apclusterK" function of the \verb"apcluster" package).
A further comparison is with the variable selection methodology for Gaussian model-based clustering (Cl VarSel) presented in Section \ref{HDMBC}. This procedure is implemented by using the \verb"clustvarsel" function included in the namesake \verb"R" package \citep{clustvarsel}. 

The number of groups $G$ is always taken as known. The default settings of each algorithm are considered, except for the $K$-means which run with 5 starts.
 As previously discussed, the RPE Clu algorithm is performed with $B=1000$, $B^*=100$, $G=\{2,4\}$ and $d=\{[10\log{(2)}]~+~1=8, \ [10\log{(4)}]+1=15\}$, respectively.

\begin{table}
  \centering
   \caption{Summary description of the simulation settings 1 - 20. When only one value for $\tau$ is given, it means that \textit{homoscedastic} Gaussian components are considered. The $*$ indicates \textit{rotated} components. }\label{tab:settings}
  \begin{tabular}{lccc}
    \toprule
    Setting	& $p$ 		& $G$ 		& $\tau$ 	 \\  \midrule
    1			&100		&2					&0.1                          \\
	2			&500		&2					&0.1                           \\
	3			&1000		&2					&0.1                             \\
	4			&100		&4					&0.1                           \\
	5			&500		&4					&0.1                            \\
	6			&1000		&4					&0.1                            \\
	7			&100		&2					&0.4                            \\
	8			&500		&2					&0.4                            \\
	9			&1000		&2					&0.4       	                  \\
	10			&100		&4					&0.4                            \\
	11			&500		&4					&0.4                            \\
	12			&1000		&4					&0.4                              \\
	13			&100		&2					&0.1-0.6\phantom{*}                          \\
	14			&100		&2					&0.1-0.3\phantom{*}                          \\
	15			&500		&2					&0.1-0.6\phantom{*}                          \\
	16			&500		&2					&0.1-0.3\phantom{*}                           \\
	17			&100		&2					&0.1-0.6*                          \\
	18			&100		&2					&0.1-0.3*                          \\
	19			&500		&2					&0.1-0.6*                          \\
	20			&500		&2					&0.1-0.3*                           \\ \bottomrule
  \end{tabular}
\end{table}

\begin{table}
  \centering
   \caption{Summary description of the simulation settings 21 - 26. Non-Gaussian distributions. }\label{tab:settings2}
  \begin{tabular}{lccl}
    \toprule
    Setting	& $p$ 	& $G$ 		& Transformation 	 \\  \midrule
	21			&100		&2			& Exp(Gaussian)     \\
	22			&100		&2			& Log(abs(Gaussian))  \\
	23			&100		&2			& Sqrt(abs(Gaussian))                       \\
	24			&100		&4			& Exp(Gaussian)     \\
	25			&100		&4			& Log(abs(Gaussian))  \\
	26			&100		&4			& Sqrt(abs(Gaussian))                       \\
 \bottomrule
  \end{tabular}
\end{table}

\begin{figure}
\centering
\includegraphics[width=0.95\textwidth]{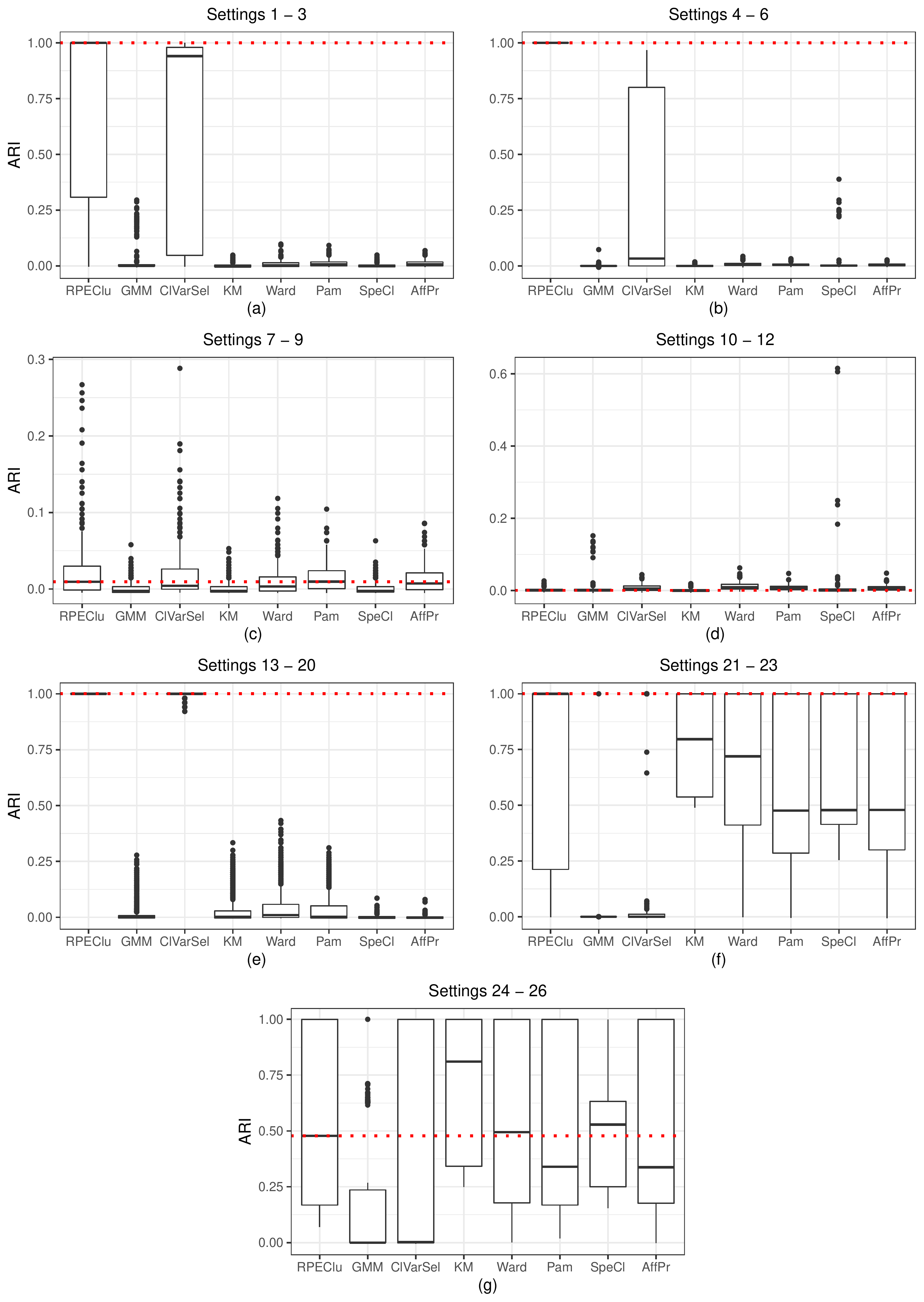}
\caption{\label{fig:sim_plots}Performance of different clustering algorithms. The labels along the horizontal axis refer to the different methods: \emph{RPEClu}, Random Projections Ensemble Clustering; \emph{GMM}, Gaussian Mixture Model; \emph{ClVarSel}, Gaussian Mixture Model with Variable Selection; \emph{KM}, $k$-means clustering; \emph{Ward}, hierarchical clustering with Ward's method; \emph{KM}, $k$-means clustering; \emph{Pam}, Partition Around Medoids algorithm; \emph{SpeCl}, spectral clustering; \emph{AffPr}, affinity propagation. The seven panels show the distribution of the Adjusted Rand Index for (a) homoscedastic Gaussian clusters ($G=2$) with highly correlated features, (b) homoscedastic Gaussian clusters ($G=4$) with highly correlated features, (c) homoscedastic Gaussian clusters ($G=2$) with mildly correlated features, (d) homoscedastic Gaussian clusters ($G=4$) with mildly correlated features (e) heteroscedastic Gaussian clusters ($G=2$), (f) non-Gaussian clusters ($G=2$) with 50\% of relevant features and (g) non-Gaussian clusters ($G=4$) with 50\% of relevant features.}
\end{figure}

\medskip
Figure \ref{fig:sim_plots} contains the aggregated results for the considered scenarios: (a)-(b) \textit{homoscedastic} Gaussian components with highly correlated features, with two and four groups respectively; (c)-(d) \textit{homoscedastic} Gaussian components with mildly related features, with two and four groups respectively; (e) \textit{heteroscedastic} Gaussian components and \textit{heteroscedastic rotated} Gaussian components; (f)-(g) \textit{non-Gaussian} components, with two and four groups respectively. The boxplots show the distribution of the ARI over $100$ simulations of each setting; the horizontal line helps the comparison with the other approaches, by highlighting the median ARI for the random projection ensemble clustering algorithm. Individual Adjusted Rand Indexes of each setting are reported in the Supplementary Material.

Results coming from this numerical study clearly show the general effectiveness of the algorithm we introduce. In fact, for all the situations considered in the boxplots of Figure \ref{fig:sim_plots}, the RPE Clu produces better solutions than those from the other state-of-the-art methods, including the two procedures that usually work well in high-dimensional contexts (i.e. spectral and affinity propagation clustering algorithms). Not surprisingly, this aspect is particularly evident in those scenarios where the original features are strongly related as some approaches tend to discard this kind of information. With reference to the Mixtures of Gaussians, for example, when $p$ is very large compared to $n$, \verb"mclust" is able to estimate only those models that have a small number of parameters, i.e. models with spherical, diagonal, or homoscedastic covariance matrix. Furthermore, the $K$-means algorithm can be viewed as a procedure which attempts to model the data as a mixture of Gaussian distributions with diagonal covariance matrices and thus it does not account for the variable correlation. Scenarios with mildly related features, i.e. 7-9 and 10-12, appear to be very hard tasks: basically all the considered methods perform poorly in terms of recovering the `true' grouping structure.

As expected, $K$-means algorithm, hierarchical agglomerative clustering with Ward's method and pam often fail because the distance measures they rely on become increasingly meaningless in high-dimensions; however, with non-Gaussian data they exhibit an acceptable performance.

A special mention should be made for the variable selection procedure (ClVarSel) that seems capable to correctly identify relevant clustering information in most of the settings. Nevertheless, it underperforms the RPE Clu, especially in the case of homoscedastic components with highly correlated features or in case of non-Gaussian data. This outcome corroborates our initial idea that feature extraction techniques are generally more effective than feature selection ones.

Globally, the capability of the RPE Clu in recovering the cluster membership does not change too much with $p$ nor with the number of groups. In addition, it is quite robust to deviations from Gaussianity: plots (f) and (g) show that RPE Clu outperforms the other methods almost always.

\section{Real data examples}
For illustration, we evaluate the performances of the clustering algorithms described in the previous section on two different real data experiments. Namely, we use the set of near infrared spectroscopic meat data originally described in the study of \cite{downey2000species} and the Lymphoma Gene Expression dataset used by \cite{chung2010sparse}.

\subsection{Meat Data} \label{subsec:meat data}
\begin{figure}
  \centering
  \includegraphics[width=\textwidth]{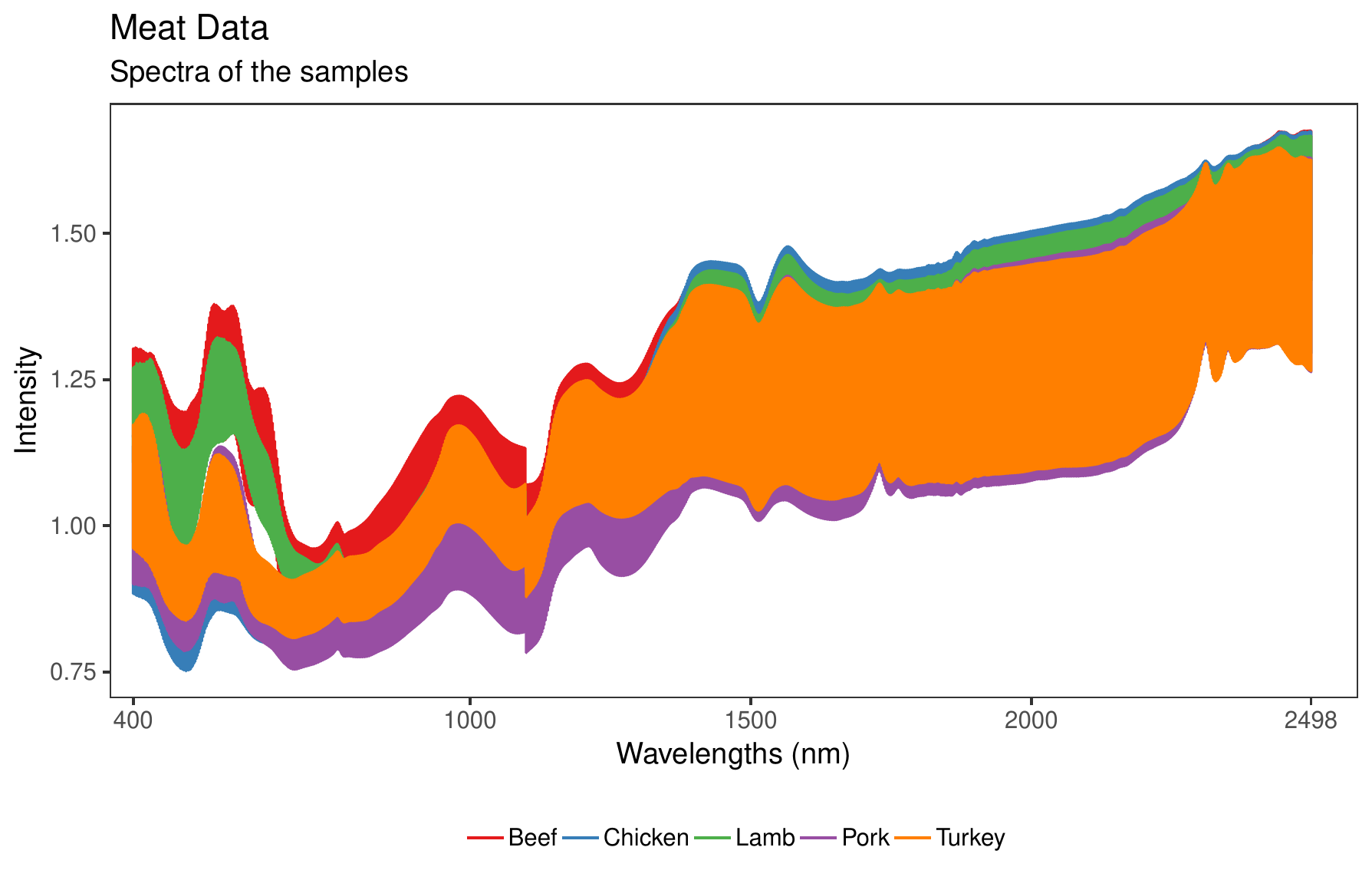}
  \caption{Meat data. Spectra of the samples, grouped by type of meat.}\label{fig:meat_spectra}
\end{figure}

This dataset contains $n=231$ samples of homogenized raw meat coming from $G=5$ different animal species. The distribution of the samples is described in Table \ref{tab:meat_distrib}. The spectra are recorded over the wavelength range 400 -- 2498 nm, with measurements taken every 2 nm. The total number of variables is thus $p=1050$.
Figure \ref{fig:meat_spectra} shows the spectrum of each sample, grouped by type of meat.

\begin{table}
  \centering
  \caption{Distribution of the meat samples}\label{tab:meat_distrib}
  \begin{tabular}{rc}
  \toprule
    Species & Samples \\ \midrule
    Beef & 32 \\
    Chicken & 55 \\
    Lamb & 34 \\
    Pork & 55 \\
    Turkey & 55 \\
    \bottomrule
  \end{tabular}
\end{table}

The objective of the analysis is to partition the set of 231 samples so as to reflect the corresponding type of meat by employing the information coming from their spectra.
The number of groups $G=5$ is taken as known; we set $B=1000$, $B^*=100$ and $d=[10\log(5)]+1=17$. Table \ref{tab:meat_res} contains the Adjusted Rand Index yielded by each method.



\begin{table}
  \centering
  \caption{ARI for the Meat Data.}\label{tab:meat_res}
  \begin{tabular}{lcr} \toprule
   Method 			& ARI 	 & Time\\ \midrule
RPEClu & 0.32 & 6460.76 \\
  GMM & 0.14 & 1.39 \\
  Clust VarSel & - & -\phantom{.1} \\
  $k$-means & 0.18 & 0.10 \\
  h-ward & 0.23 & 0.08 \\
  am & 0.18 & 0.12 \\
  Specc & 0.25 & 1.23 \\
  AClust	 & 0.18 & 1.11 \\ \bottomrule
  \end{tabular}
\end{table}

Globally, none of the employed methods is able to perfectly recover the original cluster membership of the meat data. Nevertheless, the RPE Clu algorithm provides an Adjusted Rand Index that is considerably superior to all the other solutions.
The GMM performs poorly; this is probably due to the fact that, as $p$ is very large, \verb"Mclust" could only estimate mixtures of Gaussians with spherical or diagonal covariance matrices, while data require a model that accounts for the high correlation between the features. The Clust VarSel methodology could not run because the too much correlated variables induced a model perfectly correlated with the response.

\subsection{Gene Expression Data}
The lymphoma dataset (taken from the \verb"R" package \verb"spls") contains the expression levels of $p = 4026$ genes for $n = 62$ patients. The study reports that 42 subjects have diffuse large B-cell lymphoma (DLBCL), 9 follicular lymphoma (FL), and 11 chronic lymphocytic leukemia (CLL). All gene expression profiles were base 10 log-transformed and, in order to prevent single arrays from dominating the analysis, standardized to zero mean and unit variance, as described in \cite{dettling2002supervised} and \cite{dettling2004bagboosting}.

The objective of the analysis is to group patients according to the corresponding lymphoma diagnosis, by using the information on their gene expression levels.
RPE Clu procedure run with $B=1000$, $B^*=100$ and $d=[10 \log{(3)}]+1=12$; the number of groups is taken as known and set equal to 3 for all the methods. Clustering results in terms of ARI are reported in Table \ref{tab:lymphoma_res}. As it can be seen, the performance of the random projection ensemble clustering algorithm is capable to perfectly detect the grouping structure identified by the diagnosis. Mixture of Gaussians, $K$-means and hierarchical agglomerative clustering with Ward's method provide exactly the same (good) result, up to a label switching. This is due to the fact that, when $p \gg n$, \verb"Mclust" only works on the restricted set of parsimonious models (e.g. spherical or diagonal models) and, therefore, its optimal solution often slightly improves the one yielded by the hierarchical algorithm. The mixture of Gaussians fitted to the set of 10 variables returned by Clust VarSel does not improve over the full set solution.

\begin{figure}
  \centering
  \includegraphics[width=\textwidth]{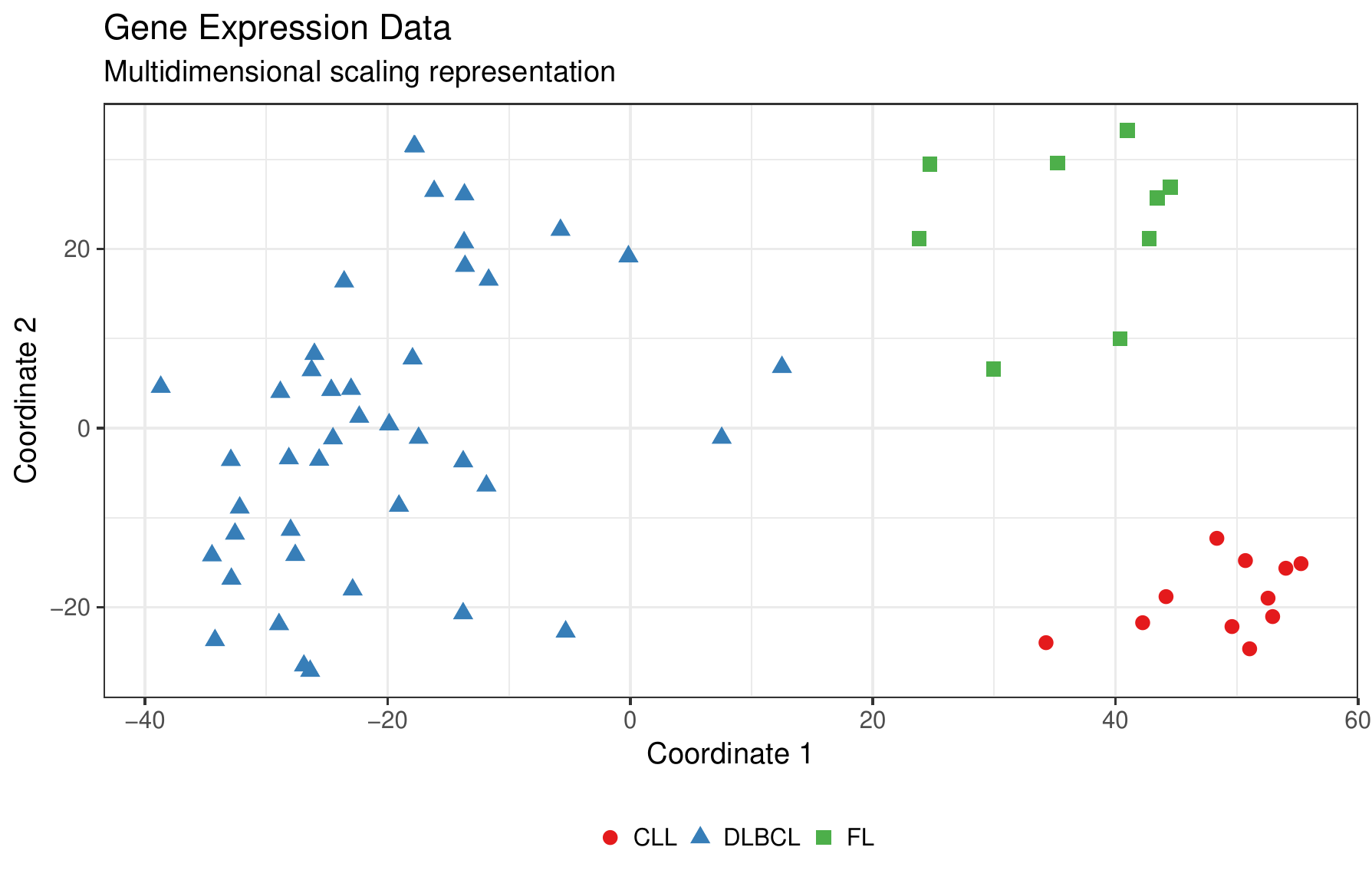}
  \caption{Gene Expression Data. Multidimensional scaling representation of the samples, grouped by type of diagnosis.}\label{fig:lymph_plot}
\end{figure}


 \begin{table}
  \centering
  \caption{ARI  for the Gene Expression Data.}\label{tab:lymphoma_res}
  \begin{tabular}{lcr}
    \toprule
    Method 			& ARI 	& Time \\ \midrule
 RPEClu  & 1.00 & 83596.81 \\
  GMM & 0.95 & 5.69 \\
  Clust VarSel & 0.44 & 5290.78 \\
  $k$-means & 0.95 & 0.10 \\
  h-ward & 0.79 & 0.04 \\
  pam & 0.84 & 0.04 \\
  Specc & 0.95 & 0.22 \\
  AClust & 0.84 & 0.35 \\
    \bottomrule
  \end{tabular}
\end{table}
\bigskip


\section{Discussion}
In this work we propose a novel procedure for model-based clustering of high-dimensional data. This procedure is based on Random Projections and it has been firstly inspired by the original idea of \cite{cannings2017random} in the context of supervised classification.

More in detail, we suggest to apply a Gaussian Mixture Model to random projections of the high-dimensional data and to select a subset of solutions accordingly to the Bayesian Information Criterion, computed here as discussed in \cite{raftery2006variable}; the multiple ‘base’ results are then aggregated via consensus  to obtain the final partition.

Such proposal has been initially motivated by some benefits associated to RPs for learning Mixture of Gaussians. \cite{dasgupta2013experiments} proved that a mixture of $G$ Gaussians can be embedded onto just $O(\log G)$ random coordinates without destroying the original group structure too much; furthermore, he demonstrated that even when the original mixing components exhibit elliptical contours, their projected counterparts are always more spherical.


Method performances, evaluated in terms of ARI with respect to the true class membership on both synthetic and real datasets, seem to confirm our motivating ideas. Overall results indeed show that RPs represent a key ingredient that decisively facilitates the learning of high-dimensional mixtures of Gaussians. Moreover, the advantage of their use in conjunction with GMM becomes even more evident as the correlation between the original variables increases. In fact, when dealing with high-dimensional sets, \verb"Mclust" search is restricted to models with few parameters only (i.e. \verb"EII", \verb"VEI", \verb"VII", \verb"VVI", \verb"EEI" and \verb"EVI") whereas data would require more complex parameterizations.

The RPE Clu algorithm is a very general tool for model-based clustering of high-dimensional data. We explore in detail its behavior within the Gaussian Mixture model framework only; however, many other distributions can in principle be used. Moreover, further options for combining the clustering results can be tested.

The number of clusters $G$ is fixed here; estimating its value is left to future work.

\bigskip
{\footnotesize \textbf{Acknowledgements.} This paper is based upon work supported by the Air Force Office of Scientific Research under award number FA9550-17-1-010.}

\bibliographystyle{Chicago}      
\bibliography{bibliography_clustering}

\end{document}